# Design and Performance Analysis of Depletion-Mode InSb Quantum-Well Field-Effect Transistor for Logic Applications

R. Islam[1], M. M. Uddin[2*], M. Mofazzal Hossain[3], M. B. Santos[4], M. A. Matin[1], and Y. Hirayama[5]

[1]Department of Electrical & Electronic Engineering, Chittagong University of Engineering & Technology (CUET), Chittagong 4349, Bangladesh

[2]Department of Physics, Chittagong University of Engineering & Technology (CUET), Chittagong 4349, Bangladesh.

[3]Department of Electronics and Communications Engineering, East West University, Plot No-A/2, Jahurul Islam City, Aftabnagar Main Road, Dhaka 1219

[4]Homer L. Dodge Department of Physics and Astronomy, University of Oklahoma, 440 West Brooks, Norman, Oklahoma 73019-2061, USA

[5]Department of Physics, Tohoku University, Sendai, Miyagi 980-8578, Japan

[*]E-mail: mohi@cuet.ac.bd

**Abstract:**

The design of a 1 µm gate length depletion-mode InSb quantum-well field-effect transistor (QWFET) with a 10 nm-thick $Al_2O_3$ gate dielectric has been optimized using a quantum corrected self-consistent Schrödinger-Poisson (QCSP) and two dimensional drift-diffusion model. The model predicts a very high electron mobility of 4.42 $m^2V^{-1}s^{-1}$ at $V_g$= 0V, a small pinch off gate voltage ($V_p$) of -0.25V, a maximum extrinsic transconductance ($g_m$) of ~ 4.94 S/mm and a drain current density of more than 6.04 A/mm. A short-circuit current-gain cut-off frequency ($f_T$) of 374 GHz and a maximum oscillation frequency ($f_{max}$) of 645 GHz are predicted for the device. These characteristics make the device a potential candidate for low power, high-speed logic electronic device applications.

**Introduction:**

The realization of improved high-performance transistors has become increasingly challenging due to difficult requirements for reduced power dissipation during operation and in stand-by mode, which are necessary to meet Moore's Law by scaling down Si-based complementary metal-oxide-semiconductor (CMOS) devices [1,2]. Moreover, increasing operating frequency and higher integration density make the power constraint issue even more challenging. Therefore, researchers have devoted much effort to develop high-performance transistors that use a low supply voltage.

Recently, many studies have focused on high-mobility channel materials such as carbon nanotubes [3], silicon nanowires [4], graphene [5], Ge [6] and III-V semiconductors [7-11]. These materials have a much higher electron mobility than in silicon. Therefore, they are potential candidates for high-speed transistors with very low supply voltage. Extensive research has been carried out with many of the III-V semiconductors (GaAs, InAs, InGaAs, GaN, GaSb, etc.) as channel material [12-17]. However, there have been fewer efforts to fabricate transistors made from InSb because of problems in the fabrication process. It is difficult to form a good interface between the top InSb (or AlInSb) layer and a subsequently deposited insulator due to a rough initial surface. There are a few reports on the formation of a high-quality gate dielectric on InSb/AlInSb and the fabrication of high-performance InSb quantum well field effect transistors (QWFETs), although the optimum layer structure for the highest performance has not been identified yet [7-10]. In this report, we determine the optimized InSb QWFET layer structure and predict its performance using a quantum corrected self-consistent Schrödinger-Poisson (QCSP) and two dimensional drift-diffusion model. It is noteworthy that our optimized InSb QWFET is expected to exhibit a very high electron mobility ($\mu$), a low pinch off gate voltage ($V_p$) and the highest cut-off frequency ($f_T$) ever reported in the InSb system. Identification of an optimized InSb QWFET structure is an important step toward realizing InSb-based high-speed and low power logic electronics applications.

**Simulation Details:**

The electronic states in confined semiconductor structures such as a quantum well are controlled by size quantization where so called energy subbands are formed. The electronic subbands of the conduction band and the corresponding envelope functions can be determined by solving the Schrödinger equation self-consistently with the Poisson equation [18]. When two or more semiconductors sandwiched together, the Fermi levels align when equilibrium is reached. The bands of an individual layer tend toward their bulk positions, resulting in band bending near the interface. The electric field created by band bending assists in bring the Fermi levels into equilibrium alignment with each other. The energy eigenvalues and wave functions are calculated by incorporating the potential due to the carrier density into the Schrödinger equation.

$$-\frac{\hbar^2}{2m^*}\frac{d^2}{dx^2}\psi(x) + V(x)\psi(x) = E\psi(x) \qquad (1)$$

where m* is the effective mass, $\hbar$ is the Planck constant $h$ divided by $2\pi$, $V(x)$ is the potential, $\psi(x)$ is the wave function and $E$ is the energy.

The electrostatic potential $\phi(x)$ is then calculated by Poisson's equation;

$$\nabla \cdot [\epsilon_0 \epsilon_r(x) \nabla \phi(x)] = -\rho(x) \qquad (2)$$

where $\rho$ is the charge per unit volume, $\varepsilon_0$ is the vacuum permittivity and the tensor $\varepsilon_r$ is the material dependent dielectric constant at position $x$. The charge density distribution $\rho(x)$ within a semiconductor device is represented by

$$\rho(x) = e\left[-n(x) + p(x) + N_D^+(x) - N_A^-(x) + \rho_{fix}(x)\right] \quad (3)$$

where $e$ is the positive elementary charge, $n$ and $p$ are the electron and hole densities, $N_D^+$ and $N_A^-$ are the ionized donor and acceptor concentrations, respectively, and $\rho_{fix}$ is the fixed or volume charge densities which arise from piezo or pyroelectric charges. In a quantum well of arbitrary potential energy profile, the potential energy $V(x)$ in Equation (1) is related to the electrostatic potential $\phi$:

$$V(x) = -q\phi(x) + \Delta E_c(x) \quad (4)$$

where $\Delta E_c(x)$ is the pseudopotential energy due to the band offset at the heterointerfaces.

Equations (1) and (2) have been iteratively used to obtain self-consistent solutions of the Schrödinger and Poisson equations. The electron density distribution function $n(x) = \sum_{k=1}^{m} \psi_k^*(x)\psi_k(x)n_k$, (where $n_k$ is electron occupation for each state and $m$ is the number of subbands) has been calculated using a trial potential $V(x)$, the wave functions $\psi(x)$, and their corresponding eigenenergies, $E_k$. The electron density in each state is described by $n_k = \frac{m^*}{\pi \hbar^2} \int_{E_k}^{\infty} \frac{1}{1 + e^{(E-E_F)/kT}} dE$, (where $K$ is the Boltzmann constant and $T$ is the measuring temperature). The electrostatic potential $\phi(x)$ in equation (4) is then calculated by using the computed $n(x)$ and donor concentration $N_D(x)$. The new potential energy $V(x)$ in equation (4) has been obtained using the computed $\phi(x)$. In this way, there is a closed loop for solving the Schrödinger equation, calculating the potential due to the resulting charge distribution, adding it to the original band-edge potential, solving Schrödinger equation again, and so on until the update is below a certain limit, indicating that convergence has been reached. Thus, the band profile of a QW has been calculated using self-consistent Schrödinger and Poisson solutions.

The temperature dependent band gap $E_g$ can be calculated by $E_g(T) = E_g(T=0) - \alpha T^2/(T+\beta)$, where the Varshni parameters for InSb are $\alpha = 0.32\ meV/K$ and $\beta = 170\ K$ [19] and for the alloy $Al_xIn_{1-x}Sb$ can be calculated by $E_g(x) = E_g(0) + (2.0\ eV)x$ [20].

The characteristics of charge transport in the QWFET is of fundamental importance for electronic device applications. Charge transport in the device has been calculated using the simplest drift-diffusion model (DDM) by taking zeroth order moments of the Boltzmann Transport Equation (BTE) and adjoining the Poisson equation [21]. The drift diffusion current density expressions for electrons and holes are

$$J_n(x) = q\mu_n n(x) E(x) + qD_n \frac{dn(x)}{dx}$$

$$J_p(x) = q\mu_p p(x) E(x) - qD_p \frac{dp(x)}{dx}$$

where $\mu_n$ and $\mu_p$ are the electron and hole mobility, $D_n$ and $D_p$ are the electron and hole diffusivity. The first part of the above equations represents the drift current, while the second part represents the diffusion current.

**DEVICE STRUCTURE**

The effects of parameters for the spacer layer, barrier layers, doping concentration, QW width and $dn_s/dV_g$ on the electron density and mobility of an InSb QWFET have been calculated using QCSP solutions, and can be found elsewhere [22]. Fig. 1 shows the cross-sectional view of an InSb QWFET structure with a 10 nm-thick $Al_2O_3$ gate insulator, which has been optimized for high electron mobility and a small $V_p$. The buffer layer of the device consists of a 3μm $Al_{0.1}In_{0.9}Sb$ layer with an $Al_{0.2}In_{0.8}Sb$ interlayer, a 20 nm-thick InSb quantum well as the channel layer, a 45 nm $Al_{0.1}In_{0.9}Sb$ barrier layer with a 25 nm spacer layer and a 2 nm Si δ-doped layer with $1\times10^{12}$ cm$^{-2}$ $n$-type doping density located 25 nm above the QW channel. Indium can be used to define the source, drain and top gate metal. Details of ALD deposition of an $Al_2O_3$ gate insulator have been presented elsewhere [10,11].

**Results and discussion:**

The band profile (conduction band minimum, $E_c$, and valence band maximum, $E_v$) of the optimized InSb QWFET at different $V_g$ was calculated using the QCSP model, and is shown in Fig. 2, where $E_F$ is at 0 eV and $E_c$ at zero depth is equal to the Schottky barrier height ($\phi_B$). The fitting parameter $\phi_B$ is determined by properties of the semiconductor surface and interface states between the gate insulator $Al_2O_3$ and top $Al_{0.1}In_{0.9}Sb$ layer. The value of $\phi_B$ is equal to the energy difference between $E_c$ at zero depth and $E_F$ under thermal equilibrium conditions. The donor density of the δ-doping layer ($1 \times 10^{12}$ cm$^{-2}$) is set so that the $E_c$ (at $V_g$=0V) is located above the $E_F$ thus precluding the formation of a parallel conduction channel. The QW is the only conduction channel. It is notable that the relatively wide band gap ~ 0.42 eV [20] of $Al_{0.1}In_{0.9}Sb$ positions the $E_v$ far away from the $E_F$ at zero depth, which prevents hole accumulation at the surface. The CBM moves to higher energy at negative $V_g$, resulting in a decrease of $n_s$ in the QW (Fig. 2). At $V_g = -0.25$V, $E_c$ is lifted above the $E_F$ which confirms the complete depletion of the $n_s$ in the QW and $E_v$ at the surface is just below $E_F$ as shown in Fig. 2. The confined 2DEG in the InSb QW is completely depleted with a very small $V_g= -0.25$V [22]. This very small pinch–off voltage, $V_p$, has been achieved due to very low interface trap density ($D_{it}$) [The Gauss law gives $Q_{it} = \varepsilon_0\varepsilon_{ox}E_{ox} - \varepsilon_0\varepsilon_{sc}E_{sc}$ (electric field $E_{ox}= V_{ox}/d_{ox}$ with Al$_2$O$_3$ thickness $d_{ox} = 10$ nm, electric field $E_{sc}$ at the semiconductor surface obtained from the QCSP simulation). The $D_{it}$ has been determined using the equation $dQ_{it}/d(E_F-E_v) = -eD_{it}$,[23]] and consequently a very large gate controllability ratio of $dn_s/dV_g = \sim 5.2 \times 10^{15} m^{-2}V^{-1}$ (estimated in the range of $-0.2V \leq V_g \leq 0V$) is predicted as discussed in ref. 22.

The gate voltage ($V_g$) dependence of total mobility ($\mu = \frac{e\langle\tau\rangle}{m^*(E)}$) of the InSb QWFET has been calculated using QCSP model. Total mobility ($\mu$) is estimated according to the Matthiessen's rule, $\frac{1}{\mu_{total}} = \frac{1}{\mu_{imp}} + \frac{1}{\mu_{bg}} + \frac{1}{\mu_{acoustic}} + \frac{1}{\mu_{polar}}$, where $\mu_{imp}$, $\mu_{bg}$, $\mu_{acoustic}$, and $\mu_{polar}$ are the mobility due to the ionized impurity, ionized background impurity, acoustic phonon impurity and polar optic phonon scattering, respectively. Details of various scattering process are described elsewhere [22]. The calculated $V_g$ dependent electron mobility ($\mu$) is shown in Fig. 3. A very high electron mobility 4.42 $m^2V^{-1}s^{-1}$ at $V_g= 0$V is achieved which is at least ~180 times greater than that of Si NMOS [7].

Figure 4 shows the well behaved I-V characteristics of the optimized InSb QWFET calculated as a function of gate bias with a 10 nm-thick Al$_2$O$_3$ gate dielectric. The plot shows a clear pinch off at a gate voltage of -0.35V which is very close to $V_p$ obtained from the band profile (Fig. 2) and the $n_s$-$V_g$ plot (in ref. 22). The $V_g$ is varied from 0 to -0.35 V with a 0.05 V step. The drain current does not show noticeable hysteresis during forward and reverse gate voltage sweeping directions. It indicates that no significant mobile bulk oxide charge is present in the gate insulator thereby the density of slow interface traps is low. The maximum drain current density $I_{ds}/W_g$ at $V_g$=0V is approximately 40 µA/µm.

The gate bias dependent drain current in the saturation region of the InSb QWFET device is shown in Fig. 5 (black squares). The device shows the quasilinear relation between $I_{ds}$ vs $V_{gs}$ in the wide bias range. The transconductance represents the ability of the FET to amplify the signal and is denoted by the output/input ratio, $g_m = dI_d/dV_{gs}$. The slope of the drain current ($I_{ds}$) yields the extrinsic $g_m$ of the 1 µm gate length device and is shown in Fig. 5 (blue squares). The maximum extrinsic $g_m$ is found to be ~ 4.94 S/mm. The very high $g_m$ of the device indicates the high speed capability of the device. Since the mobility of the device is very high, the $g_m$ is also high ($g_m$ = V µ W $c_i$ /L, where V represents the drain-source voltage, µ is the mobility, W is the gate width, $c_i$ is the gate capacitance and L is the gate length of the device). It is noted that both the quasilinear $I_{ds}$ vs $V_{gs}$ and the wide $g_m$ vs $V_{gs}$ show no significant hysteresis in forward and reverse bias directions. Counting the series resistance of the device $R_s$ ~2.5 Ω mm, the theoretical extrinsic transconductance ($g_m$) $\{g_{max}= \frac{g_m}{1+R_s g_m}\}$ is 3.7 S/mm which is almost 40% off from the peak $g_m$ value of 5.6 S/mm (Fig. 5).

The cut-off frequency $f_T$ and the maximum frequency of oscillation $f_{max}$ of the device are commonly used to measure high-speed capability. The $f_T$ is defined as the frequency of unity gain, at which the small-signal input gate current is equal to the drain current of the intrinsic FET. From S-parameter measurements, the short-circuit current-gain $f_T$ and the $f_{max}$ are determined by biasing the devices at $V_{ds}$=0.5V and $V_g$=0V. Under these conditions, the values of $f_T$ and $f_{max}$ are calculated to be 374 and 645 GHz, respectively. Extrapolating the short-circuit current gain (H$_{21}$) and the unilateral power gain (U) curves to unity and the 20 dB/decade slopes are used to determine the values of $f_T$ and $f_{max}$, respectively as shown in Fig. 6.

**Conclusions:**

We have demonstrated design and performance analysis of a depletion mode optimized InSb QWFET using a quantum corrected self-consistent Schrödinger-Poisson (QCSP) and two dimensional drift-diffusion model. The device with a 1μm gate length and 100μm gate width shows a very high electron mobility of 4.42 $m^2V^{-1}s^{-1}$ at $V_g$= 0V and a small pinch off voltage of -0.25V. The device is predicted to have the highest cut-off frequency $f_T$ and maximum frequency of oscillation $f_{max}$ ever reported in the InSb system. These results indicate that the optimized InSb QWFET has a strong potential in low power and high-speed nanoelectronics applications.

**Acknowledgements:** The authors are grateful to Dr. S. Birner for fruitful discussions.

**Figure captions**

**Figure 1.** Cross sectional view of optimized InSb QWFET layer structure with $L_G = 1\mu m$ and $W_g = 100\ \mu m$. The Si δ-doping layer is indicated by a dashed line.

**Figure 2.** Band profile of the InSb QWFET at different gate bias ($V_g$) is calculated by self-consistent Schrödinger-Poisson (SP) model. The depth is along the growth direction and the corresponding layer structures are indicated at the bottom axis. $E_c$ and $E_v$ represent the conduction-band minimum and the valence-band maximum energies, respectively and the Fermi energy $E_F$ is set to 0 eV. $E_1$ shows the first eigen energy state for different $V_g$. The vertical arrows denote the Si δ-doped regions in the AlInSb layers.

**Figure 3.** The calculated gate bias ($V_g$) dependence of electron mobility ($\mu$) in the InSb QWFET at 300 K.

**Figure 4.** Calculated drain current vs. drain bias as a function of gate bias of InSb Quantum Well FET with a 10 nm $Al_2O_3$ gate dielectric.

**Figure 5.** Calculated gate bias dependent drain current (black squares) and the extrinsic transconductance $g_m$ (blue squares) of the InSb QWFET device. The device is in the saturation region, biased at $V_{ds} = 1V$.

**Figure 6.** Calculated current ($H_{21}$) and unilateral power (U) gain as a function of frequency for an InSb QWFET device with a 1μm gate length ($L_g$) and 100 μm gate width ($W_g$). Inset: magnified view of the unity portion of the curves.

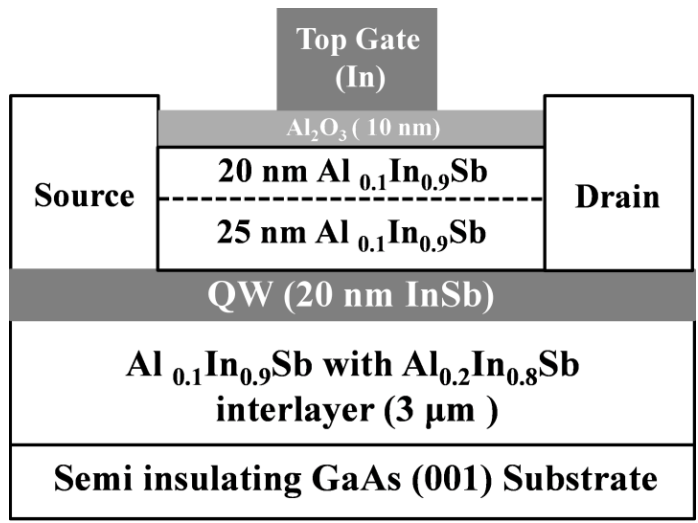

**Figure 1.**

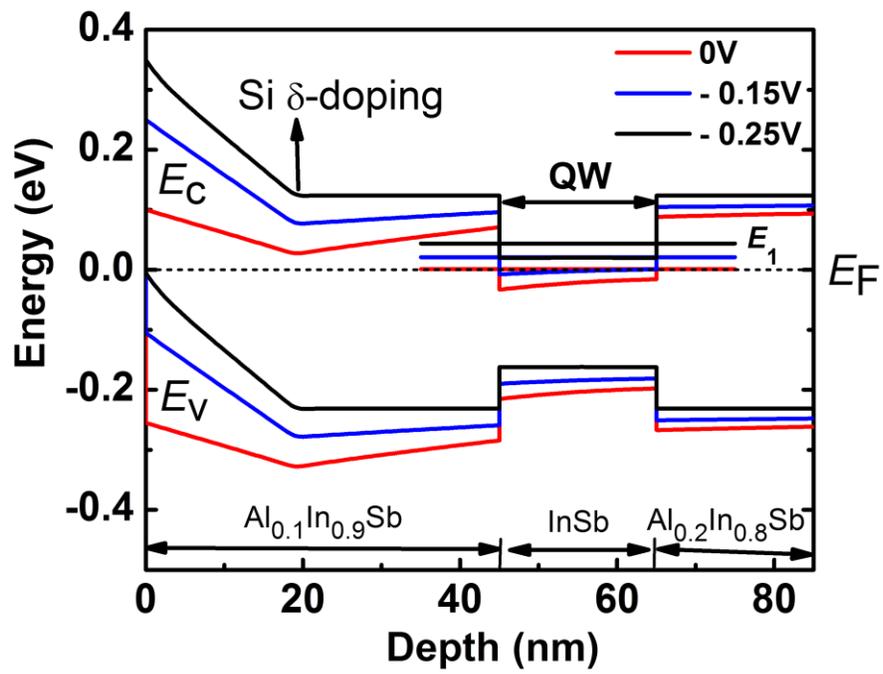

**Figure 2.**

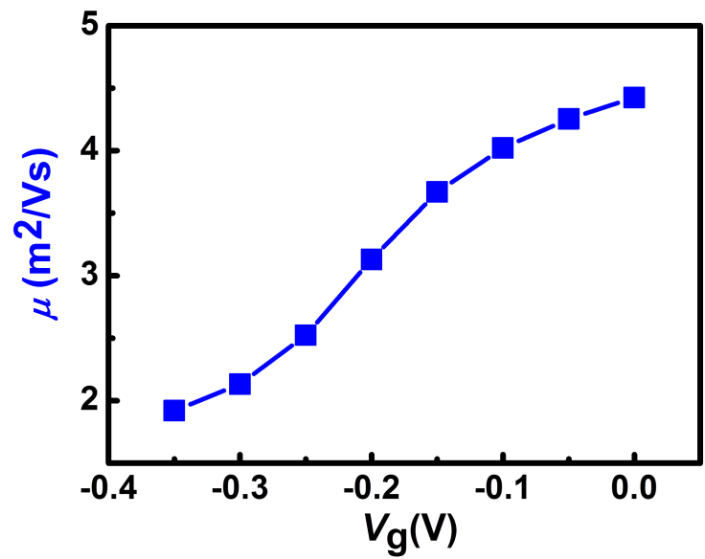

**Figure 3.**

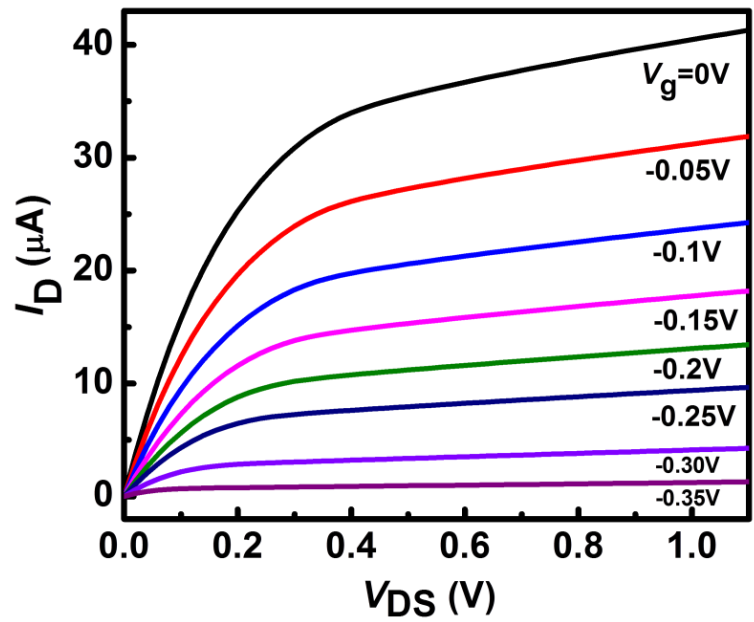

**Figure 4.**

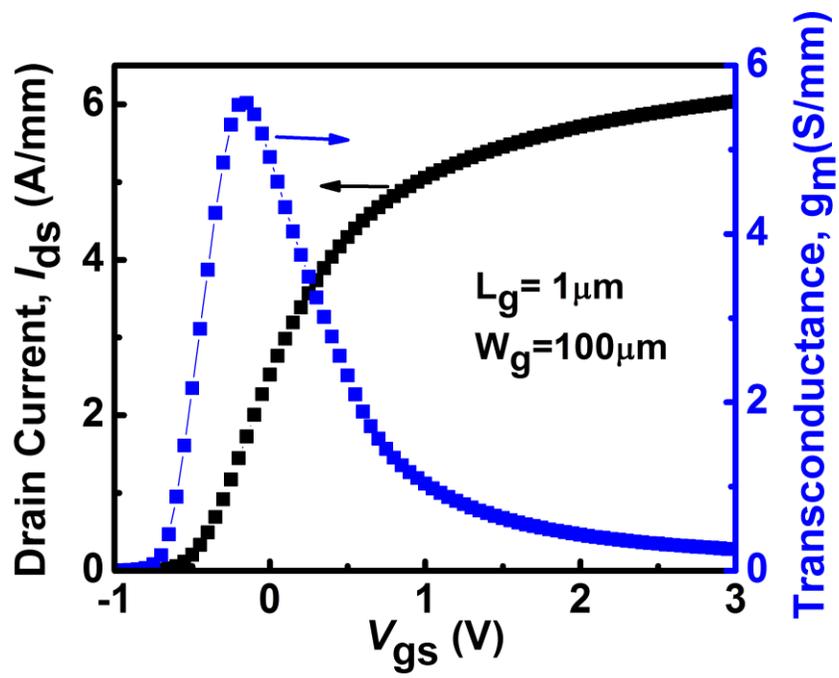

**Figure 5.**

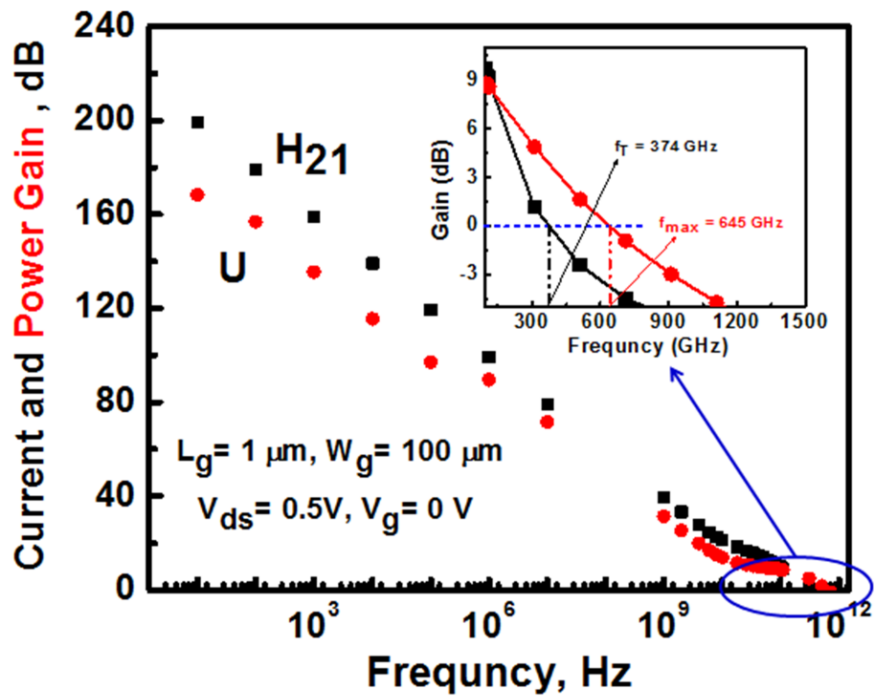

**Figure 6.**